\documentstyle[11pt]{article}
\begin{document}
\title{\bf Scattering of massive Dirac fields on the Schwarzschild black hole spacetime} 
\author{{Wei Min Jin}\\
{\small Department of Physics and Astronomy,}\\
{\small State University of New York at Buffalo,}\\
{\small Buffalo, New York 14260-1500 }}
\date{}
\font\tensl=cmsy10
\maketitle
\begin{center}
\begin{minipage}{135mm}
\vskip0.5in
\begin{center} {} \end{center}
{{\bf Abstract}. With a generally covariant equation of Dirac fields outside
a black hole, we develop a scattering theory for massive Dirac fields. The
existence of modified wave operators at infinity is shown by implementing a
time-dependent logarithmic phase shift from the free dynamics to offset a
long-range mass term. The phase shift we obtain is a matrix operator due to the existence of both positive and negative energy wave components.}
\end{minipage}
\end{center}
\vskip1in
\baselineskip15pt
\eject
\section* {1. Introduction}

As a classical scattering theory of massive Dirac fields on black hole, this paper is devoted to proving the existence of modified wave operators with logarithmic phase shifts at infinity, along the line of the treatment of the Coulomb problem by Dollard [1]. Generally speaking, the classical scattering off a black hole, on its own, does not have much physical meaning since it is essentially not observable. However, it is a first and necessary step towards the better understanding of relativistic quantum Dirac fields on black hole, which is investigated by the author elsewhere [2]. Mathematically, it is interesting since it involves some powerful mathematical techniques.    

The time-dependent scattering theory for scalar fields on the Schwarzschild
metric has been established by Dimock and Kay [3] [4] [5]
[6]. Some recent investigations of asymptotic completeness and existence have
been taken mainly by Bachelot on the Klein-Gordon equation [7] and on the
Maxwell system [8], and by Bachelot and Motet-Bachelot on the black-hole
resonance [9].    

For Dirac fields, there are some early discussions on the asymptotic behavior of Coulomb scattering by Dollard and Velo [10], and by Enss and Thaller [11].
On black hole, the existence of classical wave operators for Dirac fields has been proved at horizon (with and without mass) and at infinity (without mass) by Nicolas [12]. We proceed to
clarify the asymptotic problem at infinity for massive Dirac fields, and show that the long-range mass term in the Hamiltonian   
can actually be offset by a logarithmic phase 
shift from the free dynamics along the line of the work for massive scalar
fields by Dimock and Kay [3]. 

This paper is organized as follows. In next section, we start by showing some standard results, especially the relativistic Dirac equations on black hole, which are essential to the development of our theory.
 After acquiring enough preliminary knowledge, we develop this scattering theory of massive Dirac fields on black hole at infinity in two steps. As a first step, to get a sense of physical process, we give a rough 
qualitative picture by considering the propagation of a peaked
wave in Section 3. Then, as a second step, we use some mathematical tools to prove the existence of modified wave operators for any smooth wave in Section 4. Finally in Section 5 we add some remarks that are helpful to our understanding and future research. 

\section* {2. Preliminaries}

In the presence of gravitational field, the space-time structure is
considered as pseudo-Riemannian. However we can still fix the
locally inertial frame in which the spin representation of
Lorentz group can be established. The vierbein fields
are any collection of four vectors $V^\mu_a(x)$ which satisfy  
$$<V_a,V_b>=g_{\mu\nu}(x)V^\mu_a(x)V^\nu_b(x)=\eta_{ab},\eqno(1)$$
with a Minkowskian metric $\Bigl\{\eta_{ab}\Bigr\}=diag(1,-1,-1,-1)$.
Here Greek indices denote space-time components and Latin
indices represent spinor components. Clearly (1) is equivalent to 
$$g^{\mu\nu}(x)=V^\mu_a(x)V^\nu_b(x)\eta^{ab}.\eqno(2)$$ 
The relation 
$$g^{\mu\nu}(x)={1\over2}\Bigl\{\gamma^\mu(x),\gamma^\nu(x)\Bigr\},\eqno(3)$$
can be satisfied by choosing spinor-tensors  
$$\gamma^\mu(x)=V^\mu_a(x)\gamma^a,\eqno(4)$$
where $\gamma^a$ are Dirac matrices with the convention in [13]. The vierbein
fields $V^\mu_a(x)$ transform like coordinate vectors
and local Lorentz vectors 
$$V^\mu_a(x)=L^b_a(x)V^\mu_b(x),\eqno(5)$$
with a map $L: M\rightarrow\hbox{\tensl L}$(Lorentz group).
Then Dirac spinor fields $\psi(x): M\rightarrow C^4$, transform as
$$\psi(x)\rightarrow S(L(x))\psi(x),\eqno(6)$$
where $S$ is the finite spinor representation of the Lorentz group.  
 
The next goal is to define generally covariant derivatives that render the
transformation (6) invariant. This is achieved by the
requirement that $\gamma^\nu(x)$ have vanishing covariant derivatives [14] [15]
$$\nabla_\mu\gamma^\nu=\hbox{\tensl
D}_\mu\gamma^\nu-\Gamma_\mu\gamma^\nu+\gamma^\nu\Gamma_\mu=0.\eqno(7)$$
Then spin affine connections can be determined as [16] [2]
$$\Gamma_\mu={1\over2}G^{[a,b]}(\hbox{\tensl D}_\mu V^\nu_a)V_{\nu b},\eqno(8)$$
where $\hbox{\tensl D}_\mu$ are the covariant derivatives in space-time only, 
and the generators of the spin representation of Lorentz group are 
$$G^{[a,b]}={1\over4}[\gamma^a,\gamma^b].\eqno(9)$$
With the introduction of spin affine connections, we have the covariant
derivatives $\nabla_\mu=\partial_\mu-\Gamma_\mu$ on Dirac spinor fields
[2]. Finally, we use the Equivalence
Principle of Relativity to generalize the Dirac equation by simply replacing
regular derivatives with covariant derivatives:  
$$i\gamma^\mu(\partial_\mu-\Gamma_\mu)\Psi=m\Psi.\eqno(10)$$

Now let us consider a spherical black hole with a metric
$$\Bigl\{g_{\mu\nu}\Bigr\}=diag(\xi,-\xi^{-1},-r^2,-(r\sin\theta)^2),\eqno(11)$$
where $\xi=1-2GM/r$. From (1), the vierbein fields on such a black hole can be chosen as 
$$\Bigl\{V^\mu_a\Bigr\}=diag(\xi^{-1/2},\xi^{1/2},r^{-1},(r\sin\theta)^{-1}).\eqno(12)$$ 
Then the spin affine connection matrices of a black hole are [15] [2]
$$\Gamma_\mu=({1\over4}\xi'\gamma^1\gamma^0,0,{1\over2}\xi^{1/2}\gamma^2\gamma^1,{1\over2}\xi^{1/2}\sin\theta\gamma^3\gamma^1+{1\over2}\cos\theta\gamma^3\gamma^2).\eqno(13)$$ 
Now making a substitution in spherical coordinates,  
$$\Phi=\xi^{1/4}(-g)^{1/4}\Psi,\eqno(14)$$
where $g=\det(g^{\mu\nu})=-r^4(\sin\theta)^2$, one has a nice equation 
$$i\gamma^\mu\partial_\mu\Phi=m\Phi.\eqno(15)$$
Rewrite it as follows [14] [2] 
$$i\partial_t\Phi=H\Phi,\hskip0.1in
H=\sigma_1\otimes[\sigma_1(-i\xi{\partial\over{\partial
r}})+{\xi^{1/2}\over r}K]+(\sigma_3\otimes I)m\xi^{1/2},\eqno(16)$$
where $\sigma_i$ are Pauli matrices, $I$ is a $2\times2$ unit matrix, 
$\otimes$ represents a tensor product, and $K$ is an angular operator 
$$K=-i\sigma_2{\partial\over{\partial\theta}}-i\sigma_3{1\over{\sin\theta}}{\partial\over{\partial\phi}}.\eqno(17)$$
By using the Regge-Wheeler
variable $r_*$ defined by $dr_*/dr=\xi^{-1}$, the Hamiltonian in (16) can also be written as
$$H_{r_*}=\sigma_1\otimes[\sigma_1(-i{\partial\over{\partial
r_*}})+{\xi^{1/2}\over r}K]+(\sigma_3\otimes I)m\xi^{1/2}.\eqno(18)$$
Later in Theorem 1, we will prove the self-adjointness of this Hamiltonian. 

According to Stone's theorem [17], the self-adjointness of
Hamiltonian guarantees the existence of unitary time evolution operator
$exp(-iHt)$. Due to the existence of the black hole, Dirac fields 
asymptotically approach either horizon or infinity.  
Then we expect, for some data the Dirac
fields should approach free fields with a free Dirac Hamiltonian $H_\infty$ 
at infinity in distant past
and future. This means that the following limit should exist 
$$\lim_{t\rightarrow\pm\infty}\parallel e^{-iHt}\Phi-e^{-iH_\infty
t}\Phi_\infty\parallel=0.\eqno(19)$$
However, it turns out that the above limit does not exist at infinity for
massive Dirac fields without certain modification. The possible remedy is to
add a new phase shift $\delta(t)$ to the free wave so that the following
modified limit does exist at infinity 
$$\lim_{t\rightarrow\pm\infty}\parallel e^{-iHt}\Phi-e^{-iH_\infty
t}e^{i\delta(t)}\Phi_\infty\parallel=0.\eqno(20)$$
To understand this situation better, we like to provide a physical picture before we get into tedious mathematical proof. 

\section* {3. Physical Picture}

Suppose a black hole is located at O, and at $t=0$ a
classical Dirac spinor field is concentrated around S outside black
hole. Denote the distance OS $=|{\bf x}_0|$. For 
simplicity, let us first assume this free field can be described by a peaked
wave with a peak energy $E$. This peaked wave propagates from S at $t=0$ to near the surface of a sphere with a radius $|{\bf v}t|$ at a later time $t$, where ${\bf v}={\bf
p}/E$ is the peak velocity. After a sufficiently large time, the closest
distance between the wave and the black-hole origin is $|{\bf
v}t|-|{\bf x}_0|$, and the longest distance is $|{\bf v}t|+|{\bf
x}_0|$. Then the propagated wave $e^{-iH_\infty t}\Phi_0(0,{\bf x})$ at a
later time $t$ is almost concentrated inside a certain domain 
$$|{\bf v}t|-|{\bf x}_0|\leq|{\bf x}|\leq|{\bf v}t|+|{\bf x}_0|.\eqno(21)$$

Usually, the classical wave operators from (19) are defined by 
$$W^\pm_\infty\Phi_0=s-\lim_{t\rightarrow\pm\infty}e^{iHt}e^{-iH_\infty
t}\Phi_0.\eqno(22)$$
To prove their existence, by Cook's method it suffices to show 
$$N(t)=\parallel(H-H_\infty)e^{-iH_\infty t}\Phi_0\parallel\in L^1(R_t).\eqno(23)$$
But this is frustrated by a long-range mass term since from (16)
$$H-H_\infty\rightarrow -(\sigma_3\otimes I){GMm\over r}+O(r^{-2}).\eqno(24)$$ 
By the domain of the propagated wave (21), it suggests a lower bound for 
$$N(t)\geq{GMm\over{|{\bf v}t|+|{\bf x}_0|}}\parallel\Phi_0\parallel\rightarrow O(|t|^{-1}).\eqno(25)$$
It turns out that the lower bound of the norm $N(t)$ is of the order
$O(|t|^{-1})$, which is not integrable in $L^1(R_t)$ . So the condition (23)
is not met in the presence of long range mass term, and the classical wave
operators defined by (22) do not exist at all. 

If we can somehow offset this long range term by implementing a phase shift
from the free dynamics, we might construct some wave operators that exist at
infinity. The modified wave operators from (20) are defined by 
$$W^\pm_\infty\Phi_0=s-\lim_{t\rightarrow\pm\infty}e^{iHt}e^{-iH_\infty
t}e^{i\delta(t)}\Phi_0.\eqno(26)$$
It then suffices to show 
$$N_\delta(t)=\parallel[H-H_\infty+\delta'(t)]e^{-iH_\infty
t}e^{i\delta(t)}\Phi_0\parallel\in L^1(R_t).\eqno(27)$$
Suppose  the phase shift can be adjusted so that 
$$H-H_\infty+\delta'(t)\rightarrow ({1\over{|{\bf v}t|}}-{1\over
r})+O(r^{-2}).\eqno(28)$$
By (21), it suggests an upper bound for 
$$N_\delta(t)\rightarrow\parallel({{|{\bf x}|-|{\bf v}t|}\over{|{\bf v}t||{\bf
x}|}})e^{-iH_\infty t}e^{i\delta(t)}\Phi_0\parallel+O(|t|^{-2})$$
$$\leq{{|{\bf x}_0|}\over{|{\bf v}t|(|{\bf v}t|-|{\bf x}_0|)}}\parallel\Phi_0\parallel+O(|t|^{-2})\rightarrow O(|t|^{-2}),\eqno(29)$$
which belongs to $L^1(R_t)$. Then the condition (27) is satisfied, and the
modified wave operators defined by (26) do exist at infinity. 

By considering the propagation of a peaked wave, we have shown
roughly the nonexistence of the wave operators (22) and
the existence of the modified wave operators (26) at infinity. This gives us
a sense what is going on in this kind of scattering problems. With this
picture in mind, we now go on to provide a strict mathematical proof.  

\section* {4. Mathematical Proof} 

Generally, the classical wave operators are defined by 
$$W^\pm_\infty\Phi_\infty=s-\lim_{t\rightarrow\pm\infty}e^{iHt}\hbox{\tensl
J}e^{-iH_\infty t}\Phi_\infty,\eqno(30)$$
where {\tensl J} is a smooth bounded identifying operator from $\hbox{\tensl
H}_\infty$ to {\tensl H} to be specified later. To prove the existence of wave operators, by Cook's method [17] it suffices to show 
$$N(t)=\parallel(H\hbox{\tensl J}-\hbox{\tensl
J}H_\infty)e^{-iH_\infty t}\Phi_\infty\parallel_{\hbox{\tensl H}}\in L^1(R_t).\eqno(31)$$
As discussed previously, this is frustrated by a long-range mass term 
$$H\hbox{\tensl J}-\hbox{\tensl J}H_\infty\rightarrow -\hbox{\tensl
J}(\sigma_3\otimes I){GMm\over r}+O(r^{-2}),\eqno(32)$$ 
and the condition (31) can never be satisfied. We need to develop a
scattering mechanism to get away from the Coulomb mass term by introducing a
phase shift from free dynamics. 

Here the modified wave operators are defined by 
$$W^\pm_\infty\Phi_\infty=s-\lim_{t\rightarrow\pm\infty}e^{iHt}\hbox{\tensl
J}e^{-iH_\infty t}e^{i\delta(t)}\Phi_\infty.\eqno(33)$$
To prove their existence, it suffices to show 
$$N_\delta(t)=\parallel[H\hbox{\tensl J}-\hbox{\tensl
J}H_\infty+\overline\delta(t)]e^{-iH_\infty
t}e^{i\delta(t)}\Phi_\infty\parallel_{\hbox{\tensl H}}\in L^1(R_t),\eqno(34)$$
where 
$$\overline\delta(t)=\hbox{\tensl J}e^{-iH_\infty
t}{{d\delta(t)}\over{dt}}e^{iH_\infty t}.\eqno(35)$$
under a condition 
$$[\delta(t), \delta(s)]=0,\eqno(36)$$
for any real t and s. If we choose 
$$\overline\delta(t)=\hbox{\tensl J}(\sigma_3\otimes I){GMm\over{|{\bf
v}t|}},\eqno(37)$$
to offset the long range mass term in (31), then the phase shift
$$\delta(t)=\int^t e^{iH_\infty t'}(\sigma_3\otimes I){GMm\over{|{\bf
v}t'|}}e^{-iH_\infty t'}dt',\eqno(38)$$
does not satisfy (36), and is not self-adjoint since $[H_\infty,\sigma_3\otimes I]\not=0$. We need to find a phase shift that
commutes with $H_\infty$.

First we need to prove the self-adjointness of the
Hamiltonian (18). Let us consider a Hilbert space
$$\hbox{\tensl H}_{r_*}=L^2(R\times S^2; C^4;
dr_*d\theta d\phi)=L^2(R; dr_*)\otimes L^2(S^2; d\theta d\phi)\otimes C^4.\eqno(39)$$
Then $H_{r_*}$ is a symmetric operator on $C_0^\infty(R\times S^2;
C^4)\subset\hbox{\tensl H}_{r_*}$. We want to show it is essentially self-adjoint
(e.s.a.) on this domain along the line of [4]. 

{\bf Theorem 1}: $H_{r_*}$ is e.s.a. on $C_0^\infty(R\times S^2; C^4)\subset\hbox{\tensl H}_{r_*}$. 
 
{\bf Proof}: The Hamiltonian $H_{r_*}$ in (18) can also be written as 
$$H_{r_*}=\sigma_1\otimes H'_{r_*}+(\sigma_3\otimes I)m\xi^{1/2},$$
$$H'_{r_*}=-i{\partial\over{\partial
r_*}}\sigma_1+{\xi^{1/2}\over r}K. \eqno(40)$$
Since $\sigma_1$ is symmetric on $C^2$ and $(\sigma_3\otimes I)m\xi^{1/2}$ is
bounded, by the Kato-Rellich theorem [17] it suffices to show $H'_{r_*}$ is
e.s.a. on $C^\infty_0(R\times S^2;C^2)$ in $\hbox{\tensl
H}'_{r_*}=L^2(R\times S^2,C^2)$. The operator
$K$ of (17) is an angular operator on $L^2(S^2;C^2)$, and is e.s.a. on
$C^\infty(S^2;C^2)$, by a similar argument in [18]. Also $K$ has point spectrum of finite multiplicity: $KZ_{k}=kZ_{k},
k=0,\pm1,\pm2,...$ [14] [2], where $Z_k$ are $C^\infty$ functions. Let $Y_k$
be the eigenspace of $K$ with eigenvalues $\pm k$: $Y_k=\left\{\varphi\in
L^2(S^2;C^2): K\varphi=\pm k\varphi\right\}$.  Since
$\sigma_1K=-K\sigma_1$, we know $\sigma_1Z_k=Z_{-k}$ and hence $\sigma_1$ is symmetric in
$Y_k$. Let $\hbox{\tensl
H}_k=L^2(R)\otimes Y_k$. We define 
$$H_k=-i{\partial\over{\partial
r_*}}\otimes\sigma_1+{\xi^{1/2}\over r}\otimes kI. \eqno(41)$$
In [19], there is a proof of essential self-adjointness for the Dirac
operator on Minkowski space by Fourier 
transforming it into a multiplication operator in momentum space. This can be
applied similarly to show our radial operator
$-i\partial_{r_*}$ is e.s.a. on $C^\infty_0(R)$. Meanwhile $\sigma_1$ is symmetric and
$\xi^{1/2}r^{-1}\otimes kI$ is 
bounded, by the Kato-Rellich theorem we know $H_k$ is self-adjoint. 
Let $\hbox{\tensl H}'_{r_*}=\oplus_k\hbox{\tensl H}_k$. Then 
$H'_{r_*}=\oplus_kH_k$ is self-adjoint on 
$$D(H'_{r_*})=\Bigl\{f: f_k\in D(H_k), \sum_k\parallel
H_k f_k\parallel^2<\infty\Bigr\}.\eqno(42)$$
This extends $H'_{r_*}$ on the domain $\hbox{\tensl D}$ consisting of finite
combinations of $f\otimes Y_k$ with $f\in C_0^\infty(R)$. Now 
$C_0^\infty(R)$ is a core for $-i\partial_{r_*}$ and hence a core for $H_k$. It follows that
$\hbox{\tensl D}$ is a core for $H'_{r_*}$, namely, $H'_{r_*}$ is e.s.a. on
$\hbox{\tensl D}$. The symmetric extension of $H'_{r_*}$ to the larger domain
$C_0^\infty(R\times S^2;C^2)$ is also e.s.a. with the same closure [4]. This
completes our proof. $\Box$

Now making a new coordinate transformation 
$${dr_0\over dr}=\xi^{-1/2}(r)=(1-{2GM\over r})^{-1/2},\eqno(43)$$
we can express the new variable $r_0$ by an integral  
$$r_0(r)=\int^r_{2GM}\xi^{-1/2}(r')dr'.\eqno(44)$$
Implementing a substitution of variable 
$$r'=2GM{\cosh^2x\over\sinh^2x},\hskip0.1in\xi(r')=1-{2GM\over r'}={1\over\cosh^2x},\eqno(45)$$
we integrate by parts 
$$\int\xi^{-1/2}(r')dr'=2GM\int(\cosh x)d({\cosh^2x\over\sinh^2x})$$
$$=2GM[{\cosh^3x\over\sinh^2x}-\int{\cosh^2x\over{\cosh^2x-1}}d(\cosh x)]$$ 
$$=2GM[{\cosh^3x\over\sinh^2x}-\cosh x -{1\over2}\int({1\over{\cosh
x-1}}-{1\over{\cosh x+1}})d(\cosh x)$$
$$=2GM[{\cosh x\over\sinh^2x}+{1\over2}\log{{\cosh x+1}\over{\cosh
x-1}}].\eqno(46)$$
By (45), we write
$$r_0(r)=r\xi^{1/2}+GM\log{{1+\xi^{1/2}}\over{1-\xi^{1/2}}}.\eqno(47)$$
When $r=2GM$, we know $\xi=0$ and $r_0(2GM)=0$. When $r\rightarrow\infty$, we
have an asymptotic form 
$$r_0|_{r\rightarrow\infty}= r+GM\log(r)+O(1).\eqno(48)$$
This asymptotic form can also be derived by integrating (43)
$${dr_0\over dr}|_{r\rightarrow\infty}=\xi^{-1/2}|_{r\rightarrow\infty}\rightarrow 1+{GM\over r}+O(r^{-2}).\eqno(49)$$

By taking $dr_0/dr_*=\xi^{1/2}$, we rewrite Hamiltonian (18) as 
$$H_{r_0}=\sigma_1\otimes[\sigma_1(-i\xi^{1/2}{\partial\over\partial
r_0})+{\xi^{1/2}\over r}K]+(\sigma_3\otimes I)m\xi^{1/2},\eqno(50)$$
with a Hilbert space 
$$\hbox{\tensl H}_{r_0}=L^2((0,\infty)\times S^2; C^4;
\xi^{-1/2}dr_0d\theta d\phi).\eqno(51)$$
 
{\bf Lemma 1}: $H_{r_0}$ is e.s.a. on $C_0^\infty((0,\infty)\times S^2;
C^4)\subset\hbox{\tensl H}_{r_0}.$ 

{\bf Proof}: From the coordinate transformation $r_*\rightarrow r_0$, 
we define a unitary transformation $U_1: \hbox{\tensl
H}_{r_*}\rightarrow\hbox{\tensl H}_{r_0}$ by
$$(U_1\Phi)(r_0,\theta,\phi)=\Phi(r_*(r_0),\theta,\phi), \eqno(52)$$
This is unitary since $dr_0/dr_*=\xi^{1/2}$ and so 
$$\int(U_1\Phi)^\dagger(U_1\Phi)\xi^{-1/2}dr_0d\theta
d\phi=\int\Phi^\dagger\Phi 
dr_*d\theta d\phi.\eqno(53)$$
In general, unitary transformations preserve self-adjointness. We conclude 
$H_{r_0}$ is e.s.a. on smooth functions with compact support in $\hbox{\tensl
H}_{r_0}$ since $H_{r_*}$ is e.s.a. on smooth functions with compact support
in $\hbox{\tensl H}_{r_*}$ by Theorem 1, and since $U_1$ maps $C^\infty_0$ to
$C^\infty_0$. $\Box$

The free Hamiltonian can be written asymptotically
$r_0\rightarrow\infty$ as $r\rightarrow\infty$
$$H_\infty=\sigma_1\otimes[\sigma_1(-i{\partial\over\partial r_0})+{1\over
r_0}K]+(\sigma_3\otimes I)m,\eqno(54)$$ 
with a Hilbert space
$$\hbox{\tensl H}_\infty=L^2((0,\infty)\times S^2; C^4;
dr_0d\theta d\phi).\eqno(55)$$ 

{\bf Proposition:} $H_\infty$ is e.s.a. on $C_0^\infty((0,\infty)\times S^2;
C^4)\subset\hbox{\tensl H}_\infty$. 

{\bf Proof}: It is a special case of Lemma 1 when $M=0$. $\Box$ 

From (50) and (54) we have 
$$H_{r_0}\hbox{\tensl J}-\hbox{\tensl J}H_\infty=\hbox{\tensl
J}(\xi^{1/2}-1)H_\infty$$
$$+\hbox{\tensl J}(\sigma_1\otimes
K)\xi^{1/2}({1\over r}-{1\over r_0})-(\sigma_1\otimes\sigma_1)
(i\xi^{1/2}{d\hbox{\tensl J}\over dr_0}).\eqno(56)$$
Here the identifying operator $\hbox{\tensl J}$ is defined  
 for some $b>a>0$
$$\hbox{\tensl J}\in C^\infty(R),\hskip0.1in 0\leq\hbox{\tensl J}(r_0)\leq1,\eqno(57a)$$
$$\hbox{\tensl J}(r_0)=0\hskip0.1in for\hskip0.1in
a>r_0>0;\hskip0.1in\hbox{\tensl J}(r_0)=1\hskip0.1in for\hskip0.1in
r_0>b.\eqno(57b)$$
From (48), the second term in (56) has an asymptotic order 
$$\xi^{1/2}({1\over r}-{1\over{r_0}})={GM\log(r)\over
r^2}+O(r^{-2})= O(r^{-2+\epsilon})= O(r_0^{-2+\epsilon}),\eqno(58)$$
for $\log(r)\leq C_\epsilon r^\epsilon$ with $\epsilon>0$. The
third term in (56) vanishes as $r\rightarrow\infty$ since $d\hbox{\tensl
J}/dr_0=0$ for $r_0>b$. In the first term of (56), we use (58) to estimate
$$\xi^{1/2}-1= -{GM\over r}+O(r^{-2})= -{GM\over r_0}+O(r_0^{-2+\epsilon}).\eqno(59)$$
To offset this term, let us choose 
$$\overline\delta(t)=\hbox{\tensl J}{GM\over|{\bf v}t|}H_\infty.\eqno(60)$$
where $|{\bf v}|=\sqrt{-\triangle}/\sqrt{-\triangle+m^2}$ and $\triangle$ is
the Laplacian in spherical coordinates.  
Then from (35) we find a logarithmic phase shift
$$\delta(t)={GM\over|{\bf v}|}H_\infty\log(t),\eqno(61)$$ 
where 
$$\log(t)=sgn(t)\log|t|,\hskip0.1in {d\log(t)\over
dt}={1\over|t|}.\eqno(62)$$
By changing Hilbert space, we now work directly on the new coordinate
$r_0$ to present our main theorem.

{\bf Theorem 2}: For the classical scattering of massive Dirac fields off a
black hole characterized by a Hamiltonian: 
$$H_{r_0}=\sigma_1\otimes[\sigma_1(-i\xi^{1/2}{\partial\over{\partial r_0}})+{\xi^{1/2}\over r}K]+(\sigma_3\otimes I)m\xi^{1/2},\eqno(63)$$
with $\xi=1-2GM/r$ and 
$$K=-i\sigma_2{\partial\over{\partial\theta}}-i\sigma_3{1\over{\sin\theta}}{\partial\over{\partial\phi}},\eqno(64)$$
the modified wave operators defined by
$$W^\pm_\infty\Phi_\infty=s-\lim_{t\rightarrow\pm\infty}e^{iH_{r_0}t}\hbox{\tensl
J}e^{-iH_\infty t}e^{i\delta(t)}\Phi_\infty,\eqno(65)$$ 
do exist in $\hbox{\tensl H}_{r_0}$ by picking a time-dependent logarithmic phase shift
$$\delta(t)={GM\over|{\bf v}|}H_\infty\log(t).\eqno(66)$$  

{\bf Proof}: Since the time-dependent operator in (65) is bounded, it
suffices to show the existence of the limit in (65) for $\Phi_\infty$ in a
dense set $D_1$. $D_1$ is defined so that in Cartesian coordinates, the Fourier
transform has compact support and vanishes in a neighborhood of the
origin. Its exact definition will be 
given in Lemma 2. Then by $\epsilon/3$ arguments, we know the
existence 
of (65) for $\Phi_\infty$ in the whole Hilbert space. By Cook's method, the
existence problem is reduced to show 
$$N_\delta(t)=\parallel[H_{r_0}\hbox{\tensl J}-\hbox{\tensl
J}H_\infty+\overline\delta(t)]e^{-iH_\infty
t}e^{i\delta(t)}\Phi_\infty\parallel_{\hbox{\tensl H}_{r_0}}\in
L^1(R_t).\eqno(67)$$
To obtain this, we have to make sure that $\hbox{\tensl
J}e^{-iH_\infty t}e^{i\delta(t)}\Phi_\infty$ is in the domain of $H_{r_0}$.
 We leave this problem to Lemma 2.  

With the logarithmic phase shift (66), we may estimate the norm in (67) by
considering the asymptotic property of each term in (56)
$$N_\delta(t)\leq\parallel\hbox{\tensl J}({1\over|{\bf v}t|}-{1\over
r_0})\Phi_t\parallel_{\hbox{\tensl H}_{r_0}}+\parallel\hbox{\tensl J}Z(r_0)\Phi_t\parallel_{\hbox{\tensl H}_{r_0}},\eqno(68)$$
where $Z(r_0)$ is bounded and $Z(r_0)=O(r_0^{-2+\epsilon})$ as
$r_0\rightarrow\infty$, and also 
$$\Phi_t(t,r_0,\theta,\phi)=GMH_\infty e^{-iH_\infty
t}e^{i\delta(t)}\Phi_\infty(r_0,\theta,\phi).\eqno(69)$$
  
To change Hilbert space, we estimate 
$$\parallel\hbox{\tensl J}F\parallel_{\hbox{\tensl
H}_{r_0}}=(\int|\hbox{\tensl J}F|^2\xi^{-1/2}dr_0d\theta
d\phi)^{1/2}$$
$$\leq C(\int|\hbox{\tensl J}F|^2dr_0d\theta d\phi)^{1/2}=C\parallel\hbox{\tensl J}F\parallel_{\hbox{\tensl
H}_\infty}\leq C\parallel F\parallel_{\hbox{\tensl
H}_\infty},\eqno(70)$$
where we have used the following inequalities 
$$|\hbox{\tensl J}|^2\xi^{-1/2}\leq|\hbox{\tensl J}|^2C^2\leq C^2,\eqno(71)$$
with an upper bound $C^2=\xi^{-1/2}_{max}=\xi^{-1/2}(r_0|_{min})$. So (68) becomes
$$N_\delta(t)\leq C\parallel\hbox{\tensl J}({1\over|{\bf v}t|}-{1\over
r_0})\Phi_t\parallel_{\hbox{\tensl H}_\infty}+C\parallel Z(r_0)\Phi_t\parallel_{\hbox{\tensl H}_\infty}.\eqno(72)$$

Transforming back to Cartesian coordinates
$$x_1=r_0\sin\theta\cos\phi,\hskip0.05in x_2=r_0
\sin\theta\sin\phi,\hskip0.05in x_3=r_0\cos\theta,\eqno(73)$$
we get a matrix $\left\{u_i^j\right\}$ defined by
$dx_i=u_i^jdy_j; i,j=1,2,3$ where $y_1=r_0,y_2=\theta$ and $y_3=\phi$. Denote
the determinant $g=\det(g^{\mu\nu})=\det(g^{ij})$. To preserve the invariance
of space-time interval, it is easy to get $g({\bf x})=g({\bf y})/|\det(u)|^2$
where $\det(u)=r^2(\sin\theta)$. Then we get a transformation for half density
$$[-g({\bf x})]^{1/4}={[-g({\bf y})]^{1/4}\over|\det(u)|^{1/2}}.\eqno(74)$$
In Minkowski space-time, $\Phi_\infty=(-g)^{1/4}\Psi_\infty$, and
$(-g)^{1/4}$ transforms by (74), so we can define a
coordinate transformation operator $T(u)$ as
$$ (T(u)\Phi_\infty)({\bf
x})=({\Phi_\infty\over |\det(u)|^{1/2}})(r_0({\bf x}),\theta({\bf x}),\phi({\bf x})).\eqno(75)$$

We rewrite the free Hamiltonian from (54): 
$$H_\infty(r_0,\theta,\phi)=i\gamma^0(\gamma^1{\partial\over\partial r_0}+{\gamma^2\over
r_0}{\partial\over\partial\theta}+{\gamma^3\over{r_0\sin\theta}}{\partial\over\partial\phi})+\gamma^0m,\eqno(76)$$
and make a transformation $T(u)$ by substituting coordinates
$$TH_\infty(r_0,\theta,\phi)T^{-1}=i\gamma^0[(\gamma^1\sin\theta\cos\phi+\gamma^2\cos\theta\cos\phi-\gamma^3\sin\phi){\partial\over\partial
x_1}$$
$$+(\gamma^1\sin\theta\sin\phi+\gamma^2\cos\theta\sin\phi+\gamma^3\cos\phi){\partial\over\partial
x_2}$$
$$+(\gamma^1\cos\theta-\gamma^2\sin\theta){\partial\over\partial x_3}]+\gamma^0m=H'. \eqno(77)$$
To transform (77) into 
$$H_\infty({\bf x})=i\gamma^0(\gamma^1{\partial\over\partial
x_1}+\gamma^2{\partial\over\partial
x_2}+\gamma^3{\partial\over\partial x_3})+\gamma^0m,\eqno(78)$$
we find a spinor rotation defined by
$$S(R)\gamma^iS^{-1}(R)=R^i_j\gamma^j,\hskip0.05in
 \left\{R^i_j\right\}=\left[\begin{array}{ccc}\sin\theta\cos\phi&\sin\theta\sin\phi&\cos\theta\\
  \cos\theta\cos\phi&\cos\theta\sin\phi&-\sin\theta\\
 -\sin\phi&\cos\phi&0\end{array}\right].\eqno(79)$$
Here $R$ is orthogonal for $R^{-1}=R^T$ and $\det(R)=1$. The spinor
 representation $S(R)$ of such a rotation is unitary $S^{-1}=S^\dagger$ [13].
We may now write 
$$SH'S^{-1}=H_\infty({\bf x}).\eqno(80)$$

Let us rephrase the above problem by defining a dreibein field 
$$\Bigl\{V_i^j\Bigr\}=diag(1,r^{-1},(r\sin\theta)^{-1}). \eqno(81)$$
Then (76) can be written as
$$H_\infty(r_0,\theta,\phi)=i\gamma^0\gamma^iV_i^j\partial_j+\gamma^0m,\eqno(82)$$
where
$\left\{\partial_j\right\}=(\partial_{r_0},\partial_\theta,\partial_\phi)$.
Making a coordinate transformation
$\partial_j=(u^T)_j^k\partial'_k$, where
$\left\{\partial'_k\right\}=(\partial_{x_1},\partial_{x_2},\partial_{x_3})$,
we get the equation (77):
$$TH_\infty(r_0,\theta,\phi)T^{-1}=i\gamma^0\gamma^iV_i^j(u^T)_j^k\partial'_k+\gamma^0m=H'.\eqno(83)$$
Define a new dreibein field $(V')_i^k=V_i^j(u^T)_j^k$ in
Cartesian coordinates, and a rotation matrix $R^{-1}=\left\{(V')_i^k\right\}$
where the elements of $R$ is given by (78). Under a spinor rotation
$S(R)\gamma^iS^{-1}(R)=R_l^i\gamma^l$, we arrive at the equation (80):
$$SH'S^{-1}=i\gamma^0\gamma^lR_l^i(R^{-1})_i^k\partial'_k+\gamma^0m=i\gamma^0\gamma^l\partial'_l+\gamma^0m=H_\infty({\bf
x}).\eqno(84)$$

Finally we define a unitary transformation as a product of the
 coordinate transformation and the spinor rotation
$$U=T(u)S(R),\eqno(85a)$$ 
$$(U\Phi_\infty)({\bf
x})=({S(R)\Phi_\infty\over |\det(u)|^{1/2}})(r_0({\bf x}),\theta({\bf x}),\phi({\bf x})).\eqno(85b)$$
This operator preserves the norm squared
$$\int|(U\Phi_\infty)({\bf x})|^2d^3{\bf x}=\int|({S(R)\Phi_\infty\over|\det(u)|^{1/2}})(r_0,\theta,\phi)|^2|\det(u)|dr_0d\theta
d\phi$$
$$=\int|\Phi_\infty(r_0,\theta,\phi)|^2dr_0d\theta d\phi.\eqno(86)$$
And $U$ is unitary from $\hbox{\tensl H}_\infty$ to 
$$\hbox{\tensl H}'_\infty=L^2(R^3;C^4;d^3{\bf x}).\eqno(87)$$

Making this unitary transformation in (72) and defining $\Psi_t=U\Phi_t$, we
get 
$$N_\delta(t)\leq C\parallel\hbox{\tensl J}(|{\bf x}|)({1\over|{\bf v}t|}-{1\over
|{\bf x}|})\Psi_t\parallel_{\hbox{\tensl H}'_\infty}+C\parallel Z(|{\bf x}|)\Psi_t\parallel_{\hbox{\tensl H}'_\infty},\eqno(88)$$ 
with 
$$\Psi_t(t,{\bf x})=GMH_\infty e^{-iH_\infty t}e^{i\delta(t)}\Psi_\infty({\bf
x}),\eqno(89)$$
where $|{\bf v}|=\sqrt{-\triangle}/\sqrt{-\triangle+m^2}$ and $H_\infty=-i\mbox{\boldmath$\alpha\cdot\nabla$}+\beta m$ in Cartesian
coordinates. Note here the norms in (88) are taken on $L^2(R^3; C^4)$,
while those in (72) on $L^2(R^+\times S^2; C^4)$. 

Since $H_\infty$ has a symmetric energy
spectrum, any spinor can be expressed by a
superposition of the positive and negative energy components:
$$\Psi_\infty({\bf x})=\Psi^+_\infty({\bf x})+\Psi^-_\infty({\bf x}),\eqno(90)$$ 
where 
$$\Psi^\pm_\infty({\bf x})=\int u_\pm({\bf p})e^{\pm i{\bf p}\cdot{\bf
x}}d^3{\bf p},\eqno(91)$$ 
with spinors satisfying 
$$(\mbox{\boldmath$\alpha$}\cdot{\bf p}+\beta m)u_\pm({\bf p})=\pm E_{\bf p}u_\pm({\bf p}),\eqno(92)$$
Then the transformed wave (89) splits into two parts as well 
$$\Psi_t(t,{\bf x})=\Psi^+_t(t,{\bf x})+\Psi^-_t(t,{\bf x}),\eqno(93)$$
where
$$\Psi^\pm_t(t,{\bf x})=\pm\int GME_{\bf p}U^\pm_t({\bf p})u_\pm({\bf
p})e^{\pm i{\bf p}\cdot{\bf x}}d^3{\bf p},\eqno(94)$$ 
with two time-dependent unitary operators
$$U^\pm_t({\bf p})=\exp(\mp iE_{\bf p}t)\exp[\pm iE_{\bf p}{GM\over|{\bf
v}|}\log(t)].\eqno(95)$$
 
By the Minkowski inequality, we have
$$N_\delta(t)\leq C\parallel\hbox{\tensl J}(|{\bf x}|)({1\over|{\bf v}t|}-{1\over
|{\bf x}|})\Psi^+_t\parallel_{\hbox{\tensl H}'_\infty}+C\parallel\hbox{\tensl
J}(|{\bf x}|)({1\over|{\bf v}t|}-{1\over
|{\bf x}|})\Psi^-_t\parallel_{\hbox{\tensl H}'_\infty}$$
$$+C\parallel Z(|{\bf x}|)\Psi^+_t\parallel_{\hbox{\tensl H}'_\infty}+C\parallel
Z(|{\bf x}|)\Psi^-_t\parallel_{\hbox{\tensl H}'_\infty}.\eqno(96)$$
Since the unitary operators $U^\pm_t({\bf p})$ are of the type discussed in
[3] and each component of (94) is a smooth solution of the Klein-Gordon
equation with the Fourier transform in $C_0^\infty(R^3\backslash\left\{0\right\})$, all four terms on the right-hand side of (96) can be proved to
belong to 
$L^1(R_t)$ by the general method provided in [3]. 

Since clearly $Z(|{\bf x}|)= O(|{\bf x}|^{-2+\epsilon})\in L^2(R^3)$, the last two terms
on the right-hand side of (96) are of the order 
$$\parallel Z(|{\bf x}|)\Psi^\pm_t\parallel_{\hbox{\tensl H}'_\infty}\leq\parallel
Z(|{\bf x}|)\parallel_2\parallel\Psi^\pm_t\parallel_\infty\leq O(|t|^{-3/2+\epsilon}),\eqno(97)$$
which belong to $L^1(R_t)$ by the theorem 1(a) in [3]. By the definition of
the identifying operator {\tensl J} (57), the first two terms
on the right-hand side of (96) are of the order
$$\parallel\hbox{\tensl J}(|{\bf x}|)({1\over|{\bf v}t|}-{1\over
|{\bf x}|})\Psi^\pm_t\parallel_{\hbox{\tensl H}'_\infty}\leq O(|t|^{-2+\epsilon}),\eqno(98)$$ 
which also belong to $L^1(R_t)$ by the theorem 1(b) in [3].

With these results, we arrive at the condition (67) $N_\delta(t)\in
L^1(R_t)$, and thus prove the existence of the modified wave operators (65)
with a time-dependent logarithmic phase shift (66). As in a Corollary in [3],
it is easy to show the modified wave operators $W^\pm_\infty$ are independent
of the smooth bounded 
identifying operator {\tensl J}. $\Box$

{\bf Lemma 2}: Define a dense domain
$$D_1=\Bigl\{\phi\in\hbox{\tensl H}_\infty: (U\phi)^\sim\in
C_0^\infty(R^3\backslash\{0\};C^4)\subset\hbox{\tensl S}\Bigr\},\eqno(99)$$
where $U$ is a unitary transformation defined by (84), $(U\phi)^\sim$ is the Fourier transform of
$U\phi$, and $\hbox{\tensl S}$ is the infinitely differentiable Schwarz
space. Let $\Phi_\infty\in D_1$. Then $\hbox{\tensl
J}e^{-iH_\infty t}e^{i\delta(t)}\Phi_\infty$ is in the domain of $H_{r_0}$.

{\bf Proof}: Since $H_\infty$ does not change $D_1\subset\hbox{\tensl
H}_\infty$, with (66) we know  
$$\varphi=\exp[-iH_\infty(t-{GM\over{|{\bf v}|}}\log(t))]\Phi_\infty\in
D_1.\eqno(100)$$ 
By the previous Lemma 1, $H_{r_0}$ is e.s.a. on $C_0^\infty((0,\infty)\times S^2;
C^4)\subset D(H_{r_0})$. To show $\psi=\hbox{\tensl J}\varphi\in D(H_{r_0})$, it
suffices to show there exist  
$\psi_n\in C_0^\infty$ so that $H_{r_0}\psi_n$ converges when
$\psi_n\rightarrow\psi$. Define a series of characteristic functions $\chi_n$
for some small $a_1$ and $b_1$: $0<a_1<b_1<a$, where $a$ is the same as the
one used in defining the characteristic function {\tensl J} (57):
$$\chi_n\in C^\infty(R),\hskip0.1in 0\leq\chi_n(r_0)\leq1,\eqno(101a)$$
$$\chi_n(r_0)=0\hskip0.1in for\hskip0.1in 0<r_0<a_1\hskip0.1in or\hskip0.1in r_0>n+a_1,\eqno(101b)$$
$$\chi_n(r_0)=1\hskip0.1in for\hskip0.1in b_1<r_0<n.\eqno(101c)$$
The characteristic function $\chi_n$ cuts off wave function outside
($a_1,n+a_1$). If we define $\psi_n=\chi_n\psi\in C_0^\infty$, then we know
$\psi_n\rightarrow\chi_\infty\psi=\psi$ and 
$$H_{r_0}\psi_n=\sigma_1\otimes\sigma_1(-i\xi^{1/2})({\partial\chi_n\over\partial
r_0}\psi+\chi_n{\partial\psi\over\partial
r_0})$$
$$+[(\sigma_1\otimes K){\xi^{1/2}\over r}+(\sigma_3\otimes I)m\xi^{1/2}]\chi_n\psi$$
$$\rightarrow[\sigma_1\otimes[\sigma_1(-i\xi^{1/2}{\partial\over\partial
r_0})+{\xi^{1/2}\over r}K]+(\sigma_3\otimes I)m\xi^{1/2}]\psi,\eqno(102)$$
where we have used the result that $({\partial\chi_n}/{\partial
r_0})\psi\rightarrow0$ pointwise. Also ${\partial\chi_n}/{\partial
r_0}\leq f(r_0)\in L^2$, by the dominated convergence theorem, we know $\parallel({\partial\chi_n}/{\partial
r_0})\psi\parallel\rightarrow0$.
By (102), $H_{r_0}\psi_n$ converges to $H_{r_0}\psi$, we thus conclude $\psi\in
D(H_{r_0})$. $\Box$

\section* {5. Remarks} 

Technically, the choice of a new coordinate $r_0$ rather than
$r_*$ is of critical help to find a phase shift that commutes with the free Hamiltonian.
The time-dependent logarithmic phase shift we have derived is a
matrix operator due to the mix-up of positive and negative energy
spectrums. As a result, there is a sign difference between the phase shifts
for pure positive and negative energy waves. 

With the results of this paper about massive spinor fields and those of [3]
about massive scalar fields, we can now conclude that for the scattering of
some massive fields by black hole it is possible to find a logarithmic
phase shift from the free dynamics to offset the long range mass term at
infinity. 

If we also consider a local interaction $V(x)$, we need to replace the free
bare mass $m$ by a 
locally dressed mass $m+V(x)$ [2] in the Dirac equation. The proof of the existence of modified wave operators at
infinity in this paper should still be true as long as $V(x)\rightarrow0$
reasonably fast at infinity. 

Although we choose a simple spherical black hole as our manifold, we can also
apply our theory to any spherical stars with a Schwarzschild metric outside.

In addition to scattering, there are still many interesting problems that remain to be clarified in the relativistic theory of quantum Dirac fields on black hole such as the Hawking effect. 

\section* {Acknowledgments} 

The author would like to thank Professor Jonathan Dimock for his suggestion
of solving this problem and many valuable comments during numerous fruitful
discussions. Special thanks also go to Dr. Edward Furlani for his careful
reading and valuable comments. Finally the author would like to express 
appreciation to the referees for their reviewing this paper.  

\section* {References}

$[1]$ Dollard J 1964 {\it J. Math. Phys.} {\bf 5} 729-738\\
$[2]$ Jin W M 1998 {\it Dirac quantum fields in curved space-time} (Ph.D Thesis)\\
$[3]$ Dimock J and Kay B S 1985 {\it Class. Quan. Grav.} {\bf 3} 71-80\\
$[4]$ Dimock J 1985 {\it Gen. Rel. Grav.} {\bf 17} 353-369\\
$[5]$ Dimock J and Kay B S 1987 {\it Ann. Phys.} {\bf 175} 366-426\\
$[6]$ Dimock J and Kay B S 1986 {\it J. Math. Phys.} {\bf 27} 2520-2525\\
$[7]$ Bachelot A 1994 {\it Ann. Inst. H. Poincare Phys. Theor.} {\bf 61} no.4
411-441\\
$[8]$ Bachelot A 1991 {\it Ann.I.H.P.physique theorique} {\bf 54} 261-320\\
$[9]$ Bachelot A and Motet-Bachelot A 1993 {\it Ann.I.H.P.physique theorique} {\bf 59} 3-68\\
$[10]$ Dollard J and Velo G 1966 {\it Nuovo Cimento} {\bf 45} 801-812\\
$[11]$ Enss V and Thaller B 1986 {\it Ann. Inst. H. Poincare Phys. Theor.}
{\bf 45} no.2 

147-171\\
$[12]$ Nicolas J P 1995 {\it Ann. Inst. H. Poincare Phys. Theor.} {\bf 62} no.2
145-179\\ 
$[13]$ Bjorken J D and Drell S D 1964 {\it Relativistic Quantum Mechanics} 
(New York: 

McGraw-Hill)\\
$[14]$ Schr\"odinger E 1938 {\it Commentationes Pontif. Acad.
Sci.} {\bf 2} 321-364\\
$[15]$ Brill D R and Wheeler J A 1957 {\it Rev. Mod. Phys.} {\bf 29} 465-479\\
$[16]$ DeWitt B S 1965 {\it Dynamical theory of groups and fields} (New York: Gordon

and Breach)\\ 
$[17]$ Reed M and Simon B 1979 {\it Methods of modern mathematical physics}
Vol. III 

(New York: Academic)\\
$[18]$ Reed M and Simon B 1979 {\it Methods of modern mathematical physics}
Vol. II 

(New York: Academic)\\
$[19]$ Kato T 1976 {\it Perturbation theory for linear operators} (Berlin-New
York:  

Springer-Verlag)

\end{document}